\documentclass[twocolumn]{aastex631}

\begin{document}

\title{No Galaxy-Scale [\ion{C}{2}] Fast Outflow in the z=6.72 Red Quasar HSC J1205$-$0000}

\author[0009-0003-5438-8303]{Mahoshi Sawamura}
\affiliation{Department of Astronomy, The University of Tokyo, 7-3-1 Hongo, Bunkyo, Tokyo 113-0033, Japan}
\affiliation{National Astronomical Observatory of Japan, 2-21-1 Osawa, Mitaka, Tokyo 181-8588, Japan}

\author[0000-0001-9452-0813]{Takuma Izumi}
\affiliation{National Astronomical Observatory of Japan, 2-21-1 Osawa, Mitaka, Tokyo 181-8588, Japan}
\affiliation{Department of Astronomical Science, SOKENDAI (The Graduate University for Advanced Studies), Mitaka, Tokyo 181-8588, Japan}

\author[0000-0002-6939-0372]{Kouichiro Nakanishi}
\affiliation{National Astronomical Observatory of Japan, 2-21-1 Osawa, Mitaka, Tokyo 181-8588, Japan}
\affiliation{Department of Astronomical Science, SOKENDAI (The Graduate University for Advanced Studies), Mitaka, Tokyo 181-8588, Japan}

\author[0000-0001-7407-5329]{Takeshi Okuda}
\affiliation{Department of Astronomy, The University of Tokyo, 7-3-1 Hongo, Bunkyo, Tokyo 113-0033, Japan}
\affiliation{National Astronomical Observatory of Japan, 2-21-1 Osawa, Mitaka, Tokyo 181-8588, Japan}

\author[0000-0002-0106-7755]{Michael A. Strauss}
\affiliation{Department of Astrophysical Sciences
  Princeton University
  Princeton, NJ 08544}

\author[0000-0001-6186-8792]{Masatoshi Imanishi}
\affiliation{National Astronomical Observatory of Japan, National Institutes of Natural Sciences (NINS), 2-21-1 Osawa, Mitaka, Tokyo 181-8588, Japan}

\author[0000-0001-5063-0340]{Yoshiki Matsuoka}
\affiliation{Research Center for Space and Cosmic Evolution, Ehime University Matsuyama, Ehime 790-8577, Japan}

\author[0000-0002-3531-7863]{Yoshiki Toba}
\affiliation{National Astronomical Observatory of Japan, 2-21-1 Osawa, Mitaka, Tokyo 181-8588, Japan}
\affiliation{Academia Sinica Institute of Astronomy and Astrophysics, 11F of Astronomy-Mathematics Building, AS/NTU, No.1, Section 4,
Roosevelt Road, Taipei 10617, Taiwan}
\affiliation{Research Center for Space and Cosmic Evolution, Ehime University, 2-5 Bunkyo-cho, Matsuyama, Ehime 790-8577, Japan}

\author[0000-0003-1937-0573]{Hideki Umehata}
\affiliation{Institute for Advanced Research, Nagoya University, Furocho, Chikusa, Nagoya 464-8602, Japan}
\affiliation{Department of Physics, Graduate School of Science, Nagoya University, Furocho, Chikusa, Nagoya 464-8602, Japan}

\author[0000-0002-0898-4038]{Takuya Hashimoto}
\affiliation{Division of Physics, Faculty of Pure and Applied Sciences, University of Tsukuba, Tsukuba, Ibaraki 305-8571, Japan}
\affiliation{Tomonaga Center for the History of the Universe (TCHoU), Faculty of Pure and Applied Sciences, University of Tsukuba, Tsukuba, Ibaraki 305-8571, Japan}

\author[0000-0002-9850-6290]{Shunsuke Baba}
\affiliation{Institute of Space and Astronautical Science (ISAS), Japan Aerospace Exploration Agency (JAXA), 3-1-1 Yoshinodai, Chuo-ku, Sagamihara, Kanagawa 252-5210, Japan}

\author[0000-0002-6821-8669]{Tomotsugu Goto}
\affiliation{Institute of Astronomy, National Tsing Hua University, No.101, Section 2, Kuang-Fu Road, Hsinchu 30013,
Taiwan}

\author[0000-0002-3866-9645]{Toshihiro Kawaguchi}
\affiliation{Department of Economics, Management and Information Science Onomichi City University Hisayamada 1600-2, Onomichi, Hiroshima 722-8506, Japan}

\author[0000-0002-4052-2394]{Kotaro Kohno}
\affiliation{Institute of Astronomy, Graduate School of Science,
The University of Tokyo, 2-21-1 Osawa, Mitaka, Tokyo 181-0015, Japan}
\affiliation{Research Center for the Early Universe, Graduate School of Science, 
The University of Tokyo, 7-3-1 Hongo, Bunkyo-ku, Tokyo 113-0033, Japan}

\author[0000-0002-3848-1757]{Dragan Salak}
\affiliation{Institute for the Advancement of Higher Education, Hokkaido University, Kita 17 Nishi 8, Kita-ku, Sapporo, Hokkaido 060-0817, Japan}
\affiliation{Department of Cosmosciences, Graduate School of Science, Hokkaido University, Kita 10 Nishi 8, Kita-ku, Sapporo, Hokkaido 060-0810, Japan}

\author[0000-0002-6808-2052]{Taiki Kawamuro}
\altaffiliation{RIKEN Special Postdoctoral Researcher}
\affil{RIKEN Cluster for Pioneering Research, 2-1 Hirosawa, Wako, Saitama 351-0198, Japan}

\author[0000-0002-4923-3281]{Kazushi Iwasawa}
\affiliation{Institut de Ci\`encies del Cosmos (ICCUB), Universitat de Barcelona (IEEC-UB), Martí i Franquès, 1, 08028 Barcelona, Spain}
\affiliation{ICREA, Pg. Lluís Companys 23, 08010 Barcelona, Spain}

\author[0000-0003-2984-6803]{Masafusa Onoue}
\affiliation{Kavli Institute for the Physics and Mathematics of the Universe (WPI),The University of Tokyo Institutes for Advanced Study, The University of Tokyo, Kashiwa, Chiba 277-8583, Japan}
\affiliation{Kavli Institute for Astronomy and Astrophysics, Peking University, Beijing 100871, China}
\affiliation{Center for Data-Driven Discovery, Kavli IPMU (WPI), UTIAS, The University of Tokyo, Kashiwa, Chiba 277-8583, Japan}

\author[0000-0003-1700-5740]{Chien-Hsiu Lee}
\affiliation{W. M. Keck Observatory, 65-1120 Mamalahoa Hwy, Kamuela, HI 96743, USA}

\author[0000-0003-4814-0101]{Kianhong Lee}
\affiliation{Astronomical Institute, Tohoku University, Aramaki, Aoba-ku, Sendai, Miyagi 980-8578, Japan}
\affiliation{National Astronomical Observatory of Japan, 2-21-1 Osawa, Mitaka, Tokyo 181-8588, Japan}
\affiliation{Institute of Astronomy, Graduate School of Science, The University of Tokyo, 2-21-1 Osawa, Mitaka, Tokyo 181-0015, Japan}



\begin{abstract}
HSC 120505.09-000027.9 (J1205$-$0000) is one of the highest redshift ($z=6.72$) dust-reddened quasars (red quasars) known to date. We present an improved analysis of Atacama Large Millimeter/submillimeter Array data of the [\ion{C}{2}] $158\ \micron$ line and the underlying rest-frame far-infrared (FIR) continuum emission, previously reported in \citet{2021ApJ...908..235I}, toward J1205$-$0000. Red quasars are thought to be a transitional phase from an obscured starburst to a luminous blue quasar, in some cases associated with massive outflows driven by the active galactic nucleus (AGN). 
J1205$-$0000 has a high FIR luminosity, $L_{\mathrm{FIR}}=2.5\times 10^{12}\ L_{\odot}$ and a total IR luminosity of $L_{\mathrm{TIR}}=3.5\times 10^{12}\ L_{\odot}$, corresponding to a star formation rate (SFR) of $\sim 528\ M_{\odot}\ \mathrm{yr}^{-1}$. 
With the [\ion{C}{2}]-based dynamical mass of $\sim 1 \times 10^{11}~M_\odot$, we conclude that J1205$-$0000 is hosted by a starburst galaxy. In contradiction to \citet{2021ApJ...908..235I}, our improved analysis shows no hint of a broad component in the [\ion{C}{2}] line spectrum. Thus there is no evidence for a host galaxy-scale fast [\ion{C}{2}] outflow, despite the fact that J1205$-$0000 has fast nuclear ionized outflows seen in the rest-frame UV. 
We explore several scenarios for this discrepancy (e.g., early phase of AGN feedback, reliability of the [\ion{C}{2}] line as a tracer of outflows), and we claim that it is still too early to conclude that there is no significant negative AGN feedback on star formation in this red quasar. 
\end{abstract}



\section{Introduction} \label{sec:intro}
Mergers of gas-rich galaxies are thought to drive rapid mass growth of galaxies and supermassive black holes (SMBHs) through starburst and active galactic nuclei (AGN) phases \citep[e.g.,][]{Sanders1988, Hopkins2006, Hopkins2008}, establishing the tight observed correlations between the mass of SMBHs ($M_{\rm BH}$) and host galaxy-scale properties such as bulge mass and velocity dispersion \citep{Kormendy2013,Reines2015}. 
In this scenario, galaxy evolution starts with a merger-driven dusty starburst phase, followed by the onset of an AGN, which suppresses star formation via AGN-driven massive outflows. 
Detections of galaxy-scale AGN-driven outflows in multiphase gas \citep[e.g.,][]{2008A&A...491..407N,2012ApJ...746...86G,2014A&A...562A..21C,2016A&A...591A..28C,2020A&A...633A.134L}, 
a higher AGN fraction in interacting/merging galaxies \citep[e.g.,][]{2011ApJ...743....2S,2018PASJ...70S..37G,2018Natur.563..214K}\footnote{But see \citet{2011ApJ...726...57C, 2023OJAp....6E..34V} for a counter-argument.}, and the similar time evolutions of cosmic star formation and SMBH accretion histories \citep[][]{2014ARA&A..52..415M}, favor this picture.

It is intriguing in this context that massive (stellar mass $M_\star \sim 10^{11}~M_\odot$), quiescent old galaxies, 
are already in place at $z \sim 4-5$ \citep[e.g.,][]{2023Natur.619..716C,2023ApJ...947...20V,2024arXiv240903829W,2024NatAs.tmp....4D}, suggesting that the phase of rapid growth of galaxies and SMBHs, and associated feedback, take place at even higher redshifts.
Submillimeter (submm) studies of $z > 6$ luminous quasars \citep[1450{\AA} absolute magnitude $M_{\rm 1450} < -26$ mag, e.g.,][]{2016ApJ...816...37V,2020ApJ...904..130V,2018ApJ...854...97D} support this view, showing that their host galaxies are very massive (dynamical mass $M_{\rm dyn} \gtrsim 10^{10}~M_\odot$) with vigorous star formation (star formation rate SFR $\sim 100-1000~M_\odot$ yr$^{-1}$). 
On the other hand, some optically lower-luminous quasars ($M_{\rm 1450} > -25$ mag) found by, for example, Hyper Suprime-Cam Subaru Strategic Program (HSC-SSP) \citep{Aihara2018}, show much lower SFR, below $20~M_\odot$ yr$^{-1}$ \citep{2018PASJ...70...36I,2019PASJ...71..111I,2021ApJ...908..235I,2021ApJ...914...36I,2024arXiv240907113O}, 
implying that star formation has ceased in at least some objects at $z > 6$. 
These properties, along with the huge $M_{\rm BH}$ of these quasars \citep[$\gtrsim 10^9~M_\odot$,][]{2019ApJ...873...35S,2019ApJ...880...77O}, indeed confirm the rapid growth of galaxies and SMBHs at cosmic dawn. 

According to the above evolutionary scenario, there should be an intermediate phase between the dusty starbursts and the unobscured, optically-luminous quasars, namely {\it red quasars}. 
Red quasars have a non-negligible amount of dust extinction \citep[color excess $E(B-V) > 0.1$ mag, e.g.,][]{Richards2003,2012ApJ...757...51G,2015MNRAS.453.3932R}, but are not totally obscured: these show at least one broad optical emission line originating from the broad line region. 
Red quasars typically have high Eddington ratios and are accompanied by rest-frame UV broad absorption line (BAL) features, indicative of fast nuclear outflows \citep{Richards2003,Urrutia2009, 2015ApJ...812...66K,2018A&A...610A..31K,2023ApJ...954..156K,2024ApJ...972..171K,2024A&A...690A.283K}.
At least at $z \lesssim 3$, many red quasars have been found to host multiphase galaxy-scale outflows \citep[e.g.,][]{Brusa2018, Perrotta2019, 2016MNRAS.459.3144Z, Zakamska2019, 2022MNRAS.517.3377S, Shen2023}, exhibit high gas and dust reservoirs and elevated star formation rates \citep[e.g.,][]{2017MNRAS.465.4390B,2018MNRAS.479.1154B}.
These support the above evolutionary scenario, in which red quasars emerge from the dusty starburst by expelling the surrounding dusty medium with powerful outflows. 

However, little is known about red quasars at $z > 6$, on either nuclear or host galaxy scales. 
This is primarily due to their rarity, and their faintness in the rest-frame UV, making them difficult to identify with existing optical surveys. 
A notable case to date is the dusty starburst GNz7q at $z = 7.2$ that may host a Compton-thick low-mass quasar \citep[$M_{\rm BH} \lesssim 10^8~M_\odot$,][]{2022Natur.604..261F}, which may evolve to an unobscured quasar at a later phase. 
However, host galaxy-scale cold gas observations sufficiently sensitive to search for outflows have not yet been performed for this object.
It is therefore not clear whether the evolutionary picture proposed at low redshift also holds at cosmic dawn.

\subsection{Our target red quasar: J1205$-$0000}
As high redshift red quasars are faint at rest-frame UV wavelength, they are identified by combining deep and wide optical (or rest-frame UV) imaging survey data sensitive to the escaping quasar radiation, with infrared (IR) survey data sensitive to AGN-heated dust emission. This strategy has proven to be successful, as demonstrated in \citet{Kato2020}, who used deep $g, r, i, z, y$-band data taken by the Subaru/HSC and IR (W1 and W2 bands) survey data taken by the Wide-field Infrared Survey Explorer \citep[WISE,][]{2010AJ....140.1868W}, as part of the Subaru High-redshift Exploration of Low-Luminosity Quasars campaign \citep[SHELLQs, e.g.,][]{Matsuoka2016,2018PASJ...70S..35M,2018ApJS..237....5M}. 

\citet{Kato2020} identified two red quasars at $z > 6$, namely HSC J120505.09$-$000027.9 (hereafter J1205$-$0000) at $z_{\mathrm{MgII}} = 6.70$ and HSC J023858.09$-$031845.4 (J0238$-$0318) at $z_{\mathrm{Ly{\alpha}}} = 5.83$. In this work, we present an analysis of Atacama Large Millimeter/submillimeter Array (ALMA) observations of the $\mathrm{C^+}\ {}^{2}P_{3/2}\rightarrow{}^{2}P_{1/2}\ 157.74~\micron$ line ([\ion{C}{2}] $158~\micron$; one of the primary coolants of the cold interstellar medium) and the underlying rest-frame far-infrared (FIR) continuum emission of J1205$-$0000. J1205$-$0000 is one of the highest redshift red quasars known to date \citep{Kato2020}. 
It is relatively faint at rest-frame UV ($M_{\mathrm{1450}}=-24.4\ \mathrm{mag}$), but once its dust extinction ($E(B-V) = 0.12$ mag) is corrected for, the absolute magnitude becomes $M_{\rm 1450} = -26.1$ mag, i.e., comparable to SDSS-class high-redshift quasars. 
The extinction-corrected near-infrared (NIR) spectroscopy of the \ion{Mg}{2} line reveals that J1205$-$0000 has a BH mass of $M_{\mathrm{BH}}=2.9^{+0.3}_{-0.8}\times 10^9\ M_{\odot}$ \citep{Kato2020}. 
It is also intriguing that J1205$-$0000 shows BAL features in \ion{N}{5} and \ion{C}{4}, indicating the presence of nuclear (i.e., accretion disk-scale) outflows with a maximum velocity of $v\sim4000~\mathrm{km/s}$ \citep{2019ApJ...880...77O}. 

\citet{2021ApJ...908..235I} reported ALMA observations of [\ion{C}{2}] 158 $\micron$ and FIR continuum emission toward J1205$-$0000. They claimed detections of bright continuum emission, as well as spatially extended (out to $\sim 2''$) [\ion{C}{2}] emission that shows a complex and broad (FWHM $\sim 630$--780 km s$^{-1}$) line profile, hinting at the presence of either companion/merging galaxies or massive outflows of $\sim 100~M_\odot$ yr$^{-1}$. 
Although these results accord well with the above evolutionary scenario, we found problems with the data reduction of \citet{2021ApJ...908..235I} (see \S~\ref{sec:observation}), which resulted in a false signal of broad components in the [\ion{C}{2}] spectrum.

In this paper, we carefully re-analyse the ALMA data of J1205$-$0000 used in \citet{2021ApJ...908..235I}. Details of the data reduction, including updates from \citet{2021ApJ...908..235I}, are described in \S~\ref{sec:observation}. 
We present our main results: a careful re-assessment of the SFR and the non-detection of a [\ion{C}{2}] outflow, in \S~\ref{sec; result}. 
We provide possible interpretations of this non-detection and future prospects in \S~\ref{sec:dis}. 
Our conclusions are summarized in \S~\ref{sec:conclusion}. 
Throughout this work, we adopt the standard cosmology with $H_{0}=70\ \mathrm{km\ s}^{-1}\ \mathrm{Mpc}^{-1},\ \mathrm{\Omega_{M}}=0.3$ and $\mathrm{\Omega_{\Lambda}}=0.7$.

\section{ALMA Observations and data analysis} \label{sec:observation}
Our observations were conducted during ALMA Cycle 7 on 2020 February 26 (ID = 2019.1.00074.S; PI: T. Izumi), with the Band 6 receivers (211--275 GHz). These data are the same as used in  \citet{2021ApJ...908..235I}. The redshifted [\ion{C}{2}] 158 $\micron$ line and the underlying rest-frame FIR continuum emission were observed with 41 antennas, with baseline lengths ranging from 15.1 m to 783.5 m. The maximum recoverable scale of this observation is $\sim 6''$ and our field of view is $\sim 24''$ or $\sim128$ kpc at the distance of J1205$-$0000. J1256$-$0547 was observed as the bandpass and the flux calibrator, while J1217$-$0029 was monitored as the phase calibrator. The total on-source time was 71 minutes.

The raw data were first calibrated by using {\tt\string CASA} version 5.6 in the standard pipeline mode. This calibrated visibility data was used for the subsequent imaging step, with the {\tt\string tclean} task down to a 3$\sigma$ level, using natural weighting to maximize the point source sensitivity. As a result, we obtained a synthesized beam size of $0''.80 \times 0''.50$ (or 4.2 kpc $\times$ 2.7 kpc) with a position angle of P.A. $= -59\arcdeg$. The resultant 1$\sigma$ rms noise level is $0.29\ \mathrm{mJy\ beam}^{-1}$ at a channel spacing of $75\ \mathrm{km\ s^{-1}}$. 

\begin{figure}[th!]
\plotone{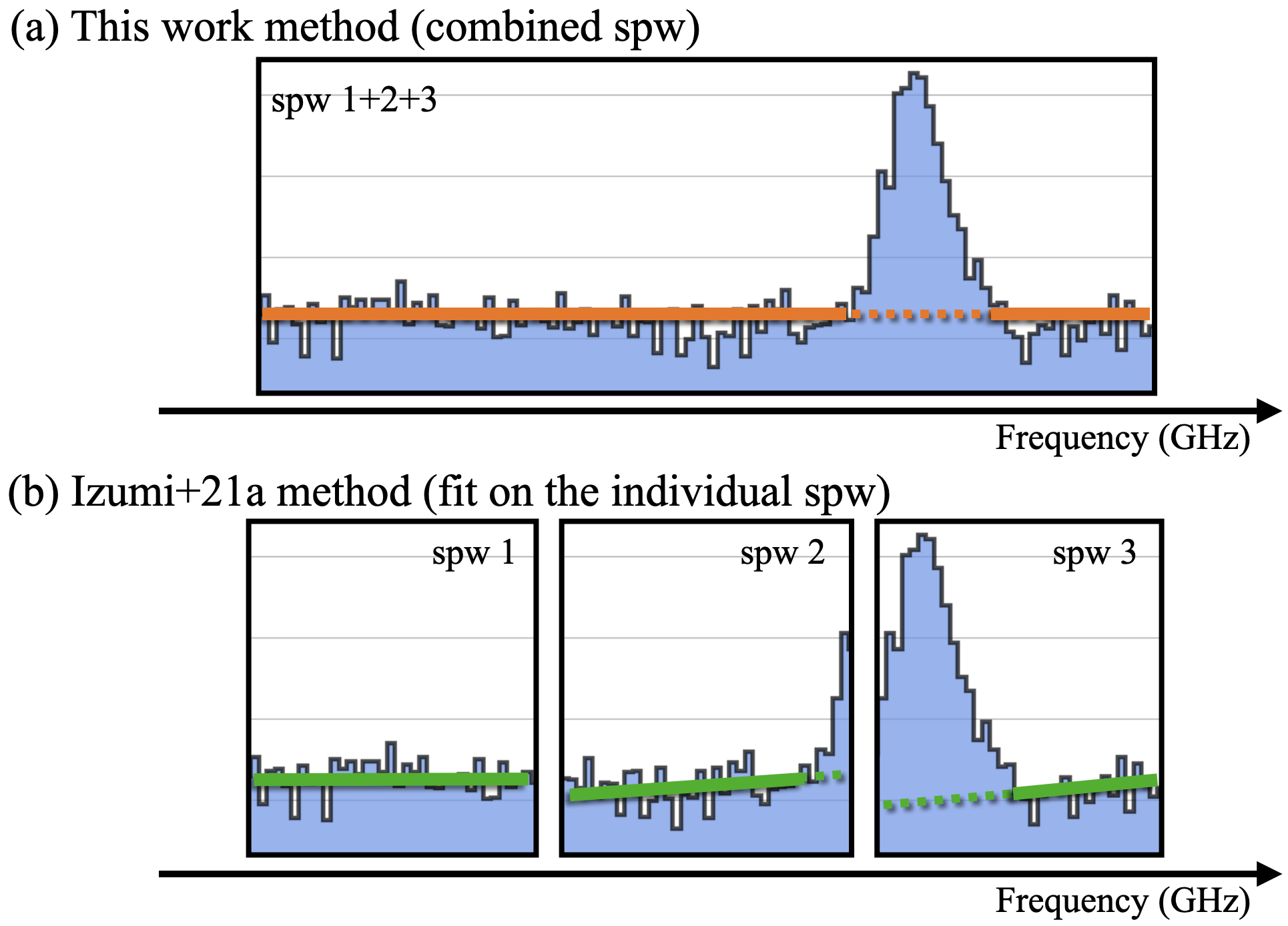}
\caption{Schematic diagram illustrating the difference between the continuum model construction methods used by \citet{2021ApJ...908..235I} and the present paper. The frequency range shown by the solid line is used for the fitting, excluding the dotted line region where the line emission is present.
(a) The method used in this study. We first combine all the spw and then model and subtract the continuum over a wide frequency range.
(b) The method used in \citet{2021ApJ...908..235I}. In this analysis, continuum is estimated and subtracted separately in each spw.
\label{fig:cartoon}}
\end{figure}

The underlying continuum emission was identified in channels free of line emission, which were subtracted from the calibrated visibilities to generate the line cube, using the {\tt\string uvcontsub} task. 
In this procedure, the continuum was modeled as a first-order polynomial, (i.e. straight line), fit over all three spectral windows (spw; each has 1.875 GHz width) simultaneously (Figure \ref{fig:cartoon}(a)).
This is the critical difference from the analysis of \citet{2021ApJ...908..235I}, who fit and subtracted the continuum emission in each of the three spw separately (Figure \ref{fig:cartoon}(b)). 
Due to the limited frequency coverage in each spw, the resultant continuum-subtracted cube had substantial uncertainties in the zero-level baseline that varied from one spw to another. 
This, together with the fact that the [\ion{C}{2}] line of this quasar falls in the overlap between two spw, generated a false broad component in the resultant [\ion{C}{2}] cube in the analysis of \citet{2021ApJ...908..235I}\footnote{Note that in other work by our team \citep{2018PASJ...70...36I,2019PASJ...71..111I,2021ApJ...914...36I}, continuum subtraction was performed after combining spw, as shown in Figure \ref{fig:cartoon}(a). Therefore, the conclusions of those papers are robust.}. 
Our current analysis fitting the continuum over the three spw simultaneously should give a more robust continuum subtraction, and thence the shape of the [\ion{C}{2}] spectrum. 
This continuum emission itself is mapped using the {\tt\string tclean} task down to the $3\sigma$ level ($1\sigma\sim18\ \mu\mathrm{Jy/beam}$). In the following, we present statistical errors only unless mentioned otherwise. According to the ALMA Cycle 7 Technical Handbook, the systematic flux calibration error for ALMA Band 6 is $\sim10\%$.

\section{Results} \label{sec; result}

\subsection{FIR continuum emission} \label{subsec; cont}
\begin{figure*}[t!]
\plotone{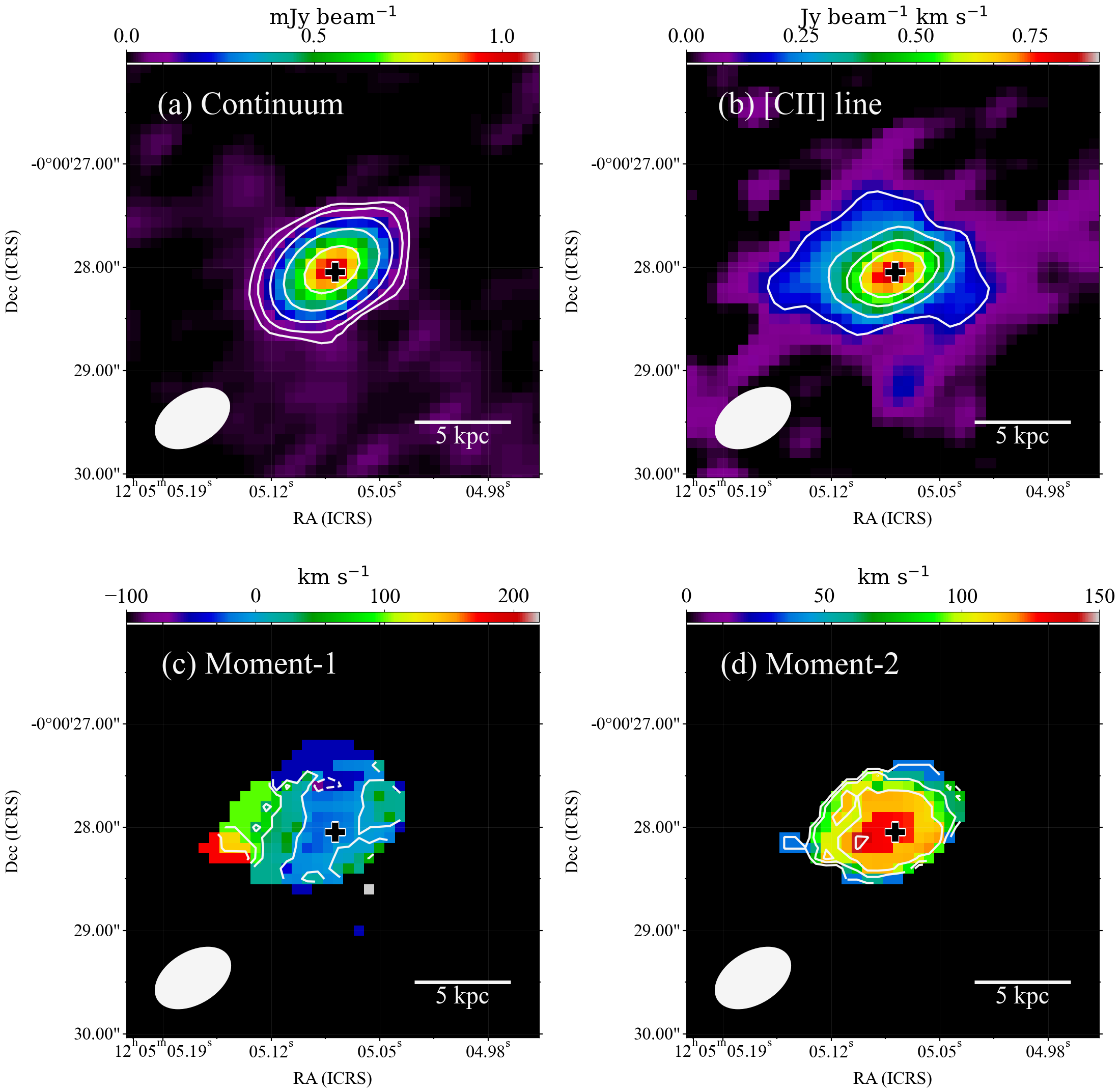}
\caption{(a) The rest-frame FIR continuum image of J1205$-$0000 with the synthesized beam of $0''.79\times 0''.49$ at the bottom left corner. Contours are shown at $3,\ 5,\ 10,\ 20,\ 40\sigma$ ($1\sigma=17.6\ \mathrm{\mu Jy\ beam}^{-1}$). (b) The [\ion{C}{2}] $158\ \micron$ line moment-0 map with the synthesized beam of $0.''80\times 0''.50$. Contours are drawn at $3,\ 6,\ 9,\ 12\sigma$ ($1\sigma=0.048\ \mathrm{Jy\ beam}^{-1}\ \mathrm{km\ s}^{-1}$). (c) Intensity-weighted [\ion{C}{2}] velocity map. Contours are overlaid at every $50\ \mathrm{km\ s}^{-1}$ from $-100$ to $200\ \mathrm{km\ s}^{-1}$. (d) Intensity-weighted velocity dispersion map. Contours represent 35, 70, 105, and 140 $\mathrm{km\ s}^{-1}$. The beam sizes for (c) and (d) are the same as (b).
In each panel the bottom-right bar indicates 5 kpc length. The black plus  represents the center of the rest-frame UV light distribution observed by HSC \citep{Matsuoka2016}.}
\label{fig:two}
\end{figure*}

\begin{deluxetable*}{c|ccc}[t!]
\tablewidth{0pt} 
\tablecaption{Observed and Derived Physical values of J1205$-$0000\label{tab:deluxesplit}}
\tablehead{
    & \multicolumn{3}{c}{Continuum Emission ($T_d = 47\ K,\ \beta=1.6,\ \kappa_{\mathrm{250GHz}}=0.4\ \mathrm{cm}^2\ \mathrm{g}^{-1}$)}
}
\startdata
$f_{1.2\mathrm{mm}}$ (mJy) & \multicolumn{3}{c}{$1.19\pm 0.02$} \\
$L_{\mathrm{FIR}}\ (10^{12}\ L_{\odot})$ & \multicolumn{3}{c}{$2.51\pm 0.04$} \\
$L_{\mathrm{TIR}}\ (10^{12}\ L_{\odot})$ & \multicolumn{3}{c}{$3.54\pm 0.05$} \\
$M_{\mathrm{dust}}\ (10^8\ M_{\odot})$ & \multicolumn{3}{c}{$1.79\pm 0.03$} \\
$\mathrm{SFR_{TIR}}\ (M_{\odot}\ \mathrm{yr}^{-1})$ & \multicolumn{3}{c}{$528\pm 8$} \\
Size\ (beam-deconvolved) & \multicolumn{3}{c}{$(0''.27\pm 0''.06)\times (0''.21\pm 0''.08)=(1.4\pm 0.3)\ \mathrm{kpc}\times (1.1\pm 0.4)\ \mathrm{kpc}$} \\
\cline{1-4}
& \multicolumn{3}{c}{[\ion{C}{2}] $158\ \micron$ Line Emission} \\
\cline{2-4}
& \multicolumn{1}{c}{This work} && \multicolumn{1}{c}{Izumi et al. 2021a} \\
\cline{1-4}
$S_{\mathrm{[CII]}}\Delta V\ (\mathrm{Jy\ km\ s}^{-1})$ & \multicolumn{1}{c}{$1.32 \pm 0.15$} && \multicolumn{1}{c}{$1.69 \pm 0.07$}\\
$L_{\mathrm{[CII]}}\ (10^{9}\ L_{\odot})$ & \multicolumn{1}{c}{$1.46\pm 0.22$} && \multicolumn{1}{c}{$1.87\pm 0.08$} \\
$\mathrm{SFR_{\mathrm{[CII]}}}\ (M_{\odot}\ \mathrm{yr}^{-1})$ & \multicolumn{1}{c}{$127\pm 19$} && \multicolumn{1}{c}{$163\pm 7$\tablenotemark{a}} \\
Size\ (beam-deconvolved) & \multicolumn{1}{c}{$(1''.02\pm 0''.13)\times (0''.56\pm 0''.12)$} && \multicolumn{1}{c}{$(1''.43\pm 0''.36)\times (0''.77\pm 0''.33)$} \\
& \multicolumn{1}{c}{$=(5.4\pm 0.6)\ \mathrm{kpc}\times (3.0\pm 0.7)\ \mathrm{kpc}$} && \multicolumn{1}{c}{$=(7.7\pm 1.9)\ \mathrm{kpc}\times (4.1\pm 1.8)\ \mathrm{kpc}$} \\
$M_{\mathrm{dyn}}\ (10^{10}\ M_{\odot})$\ (turbulent sphere) & \multicolumn{1}{c}{$2.81\pm0.83$} \\
$M_{\mathrm{dyn}}\ (10^{10}\ M_{\odot})$\ (thin rotating disk) & \multicolumn{1}{c}{$13$} \\
\cline{1-4}
& \multicolumn{1}{c}{Single Gaussian} & \multicolumn{2}{c}{Double Gaussian} \\
\cline{3-4}
&& \multicolumn{1}{c}{Core} & \multicolumn{1}{c}{Wing} \\
\cline{1-4}
$z_{\mathrm{[CII]}}$ & $6.7224\pm 0.0003$ &  &  \\
Peak flux density\ ($\mathrm{mJy}$) & $2.92\pm 0.16$ & $2.08\pm 1.95$ & $0.93\pm 2.00$ \\
$\mathrm{FWHM_{\mathrm{[CII]}}}$\ ($\mathrm{km\ s}^{-1}$) & $404\pm 27$ & $328\pm 133$ & $616\pm 407$ \\
$\chi^2/\mathrm{d.o.f}$ & $0.92$ & \multicolumn{2}{c}{$0.92$} \\
\enddata
\tablenotetext{a}{This is derived using the calibration of local HII/starburst galaxies by \citet{derooze14}, the same as our method, differs which from the approach used in \citet{2021ApJ...908..235I}.}
\end{deluxetable*}

Figure \ref{fig:two}a shows the rest-frame FIR (observed at $\lambda_{\rm obs} = 1.2$ mm) continuum emission of J1205$-$0000. 
We fit this emission distribution with a 2D Gaussian function, using the {\tt\string imfit} task in {\tt\string CASA}, 
resulting in a beam-deconvolved size of $\sim 0''.2$. 
The coordinates of the emission peak are $(\alpha_{\mathrm{ICRS}},\ \delta_{\mathrm{ICRS}})=(12^{\mathrm{h}} 05^{\mathrm{m}} 05^{\mathrm{s}}.0802,\ -00^{\circ}00'28''.0365)$, which almost coincides with the position from HSC observations of $(12^{\mathrm{h}} 05^{\mathrm{m}} 05^{\mathrm{s}}.0789,\ -00^{\circ}00'28''.0465)$. 
The peak flux density at this position is $1.01\ \mathrm{mJy\ beam}^{-1}$, resulting in a high signal-to-noise ratio of $\sim50$. 
The total continuum flux density, measured after smoothing the beamsize to $1''.5$ with the {\tt\string CASA} task {\tt\string imsmooth}, is $1.19\ \mathrm{mJy}$.

The FIR continuum emission of high redshift objects primarily originates from thermal dust heated by star formation \citep[e.g.,][]{Yun2000, Carilli2001}.
Here we estimate the FIR luminosity ($L_{\mathrm{FIR}};\ 42.5\mathrm{-}122.5\ \micron$) and total IR luminosity ($L_{\mathrm{TIR}};\ 8\mathrm{-}1000\ \micron$) by assuming an optically thin modified blackbody spectrum with a dust temperature of $T_{\mathrm{dust}}=47\ \mathrm{K}$ and emissivity index $\beta=1.6$, values that are often adopted for high-redshift optically luminous quasars  \citep[e.g.,][]{Beelen2006, 2013ApJ...772..103L}. We also correct the contrast to account for the heating effect of the cosmic microwave background (CMB) \citep{2013ApJ...766...13D}. 
With these, we estimate $L_{\mathrm{FIR}}=(2.51\pm 0.04)\times 10^{12}\ L_{\odot}$ and $L_{\mathrm{TIR}}=(3.54\pm 0.05)\times 10^{12} L_{\mathrm{\odot}}$, respectively. 
However, note that there seems to be a wide variety of dust temperatures over the range $30\mathrm{-}87~\mathrm{K}$ for luminous quasars at $z > 5$ \citep[e.g.,][]{Beelen2006, 2013ApJ...772..103L, 2022MNRAS.515.1751W, 2024arXiv240104211T}, which are not spatially resolved, resulting in $L_{\mathrm{TIR}}=(1.2\mathrm{-}8.4)\times 10^{12}$ with $\beta=1.6$.
A more accurate estimate of $T_{\rm dust}$ will be constrained by future multi-wavelength observations. 
To derive the dust mass, we adopt a rest-frame mass absorption coefficient of $\kappa_{\nu}=0.4\ (\nu/250\mathrm{GHz})^{\beta}\ \mathrm{cm}^2\ \mathrm{g}^{-1}$ \citep{Hildebrand1983, Alton2004}, resulting in $M_{\mathrm{dust}}=(1.79\pm 0.03)\times 10^8\ M_{\odot}$.


\subsection{[\ion{C}{2}] line emission} \label{subsec; line}
\begin{figure}[th!]
\plotone{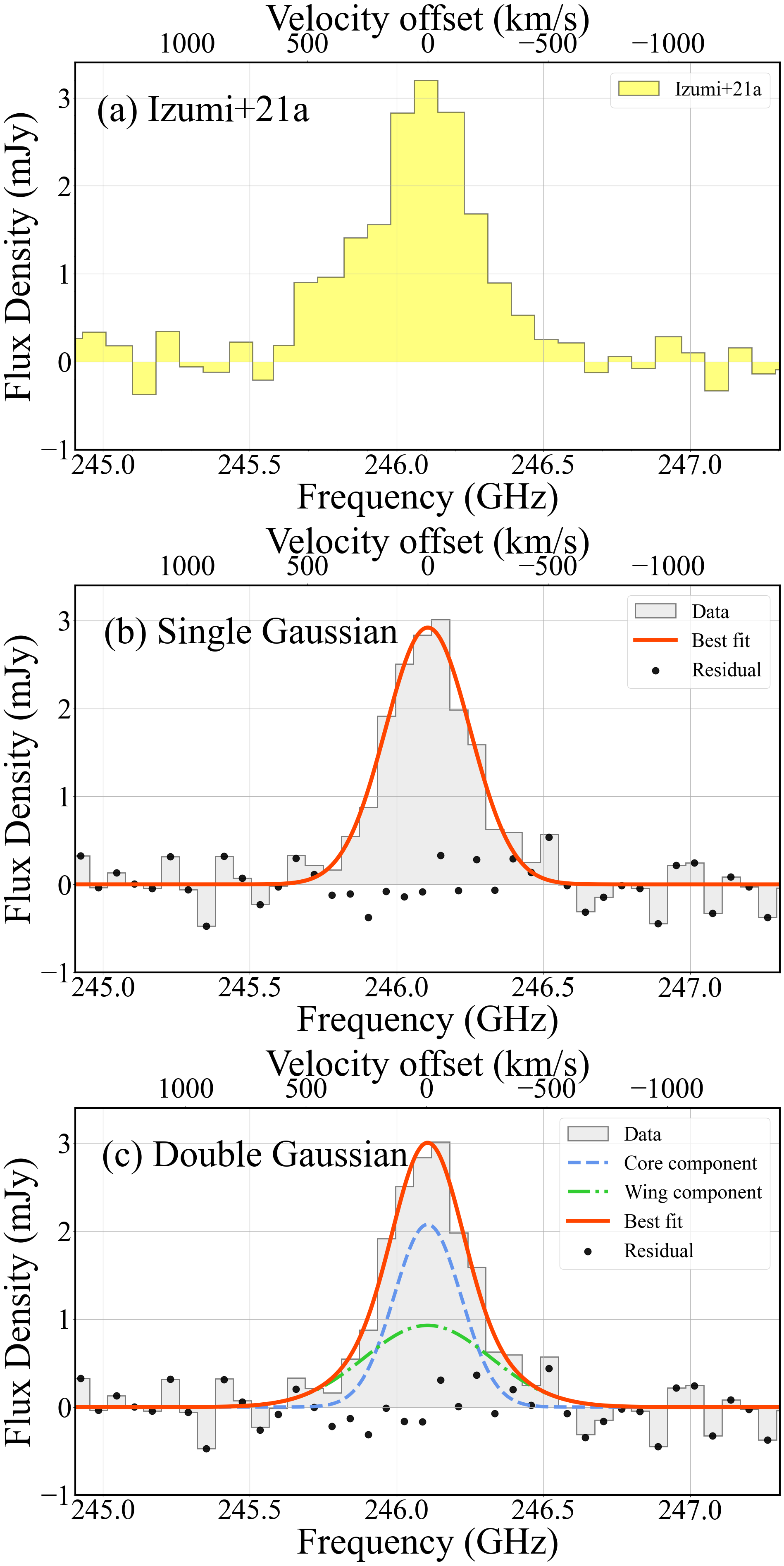}
\caption{(a) The [\ion{C}{2}] spectrum of J1205$-$0000 reported in \citet{2021ApJ...908..235I}. 
(b) [\ion{C}{2}] $158\ \micron$ line spectrum of J1205$-$0000 measured with a $1''.5$ diameter circular aperture placed at the AGN position. The spectrum is modelled with a single Gaussian function. It is evident that the complex broad component seen in (a) now disappears here. This difference is due to the improved and more robust continuum modeling in the present paper.
(c) Double Gaussian fit to the same [\ion{C}{2}] spectrum as in (b). We forcibly fit the core component and the broad wing component. The latter is used to represent possible outflows. 
The black points shown in (b) and (c) represent the residuals of each fit.
The reduced $\chi^{2}$ values for these fits are given in Table \ref{tab:deluxesplit}. The two are very similar, suggesting no statistical detection of the broad component. 
\label{fig:spectra}}
\end{figure}

We made the velocity-integrated moment-0 map by integrating over $\pm 750\ \mathrm{km\ s}^{-1}$ relative to the line center (Figure \ref{fig:two}b). This velocity range was chosen to possibly encompass the broad outflow component.
It seems that the [\ion{C}{2}] emission is spatially more extended than the continuum emission: 
this is confirmed by the 2D Gaussian fit onto this [\ion{C}{2}] map with the {\tt\string imfit} task, 
resulting in a beam-deconvolved size of $\sim 1''.0 \times 0''.6$ (Table \ref{tab:deluxesplit}). 
The total [\ion{C}{2}] line flux ($S_{\mathrm{[CII]}}\Delta V$) was measured in the moment-0 map by smoothing the beam size to $1''.5$ (enough to encompass all apparent emission) in the image plane, yielding $S_{\mathrm{[CII]}}\Delta V =1.32\pm 0.15\ \mathrm{Jy\ km\ s}^{-1}$. 
We derive the [\ion{C}{2}] line luminosity following \citet{Solomon2005}: $L_{\mathrm{[CII]}}=1.04\times 10^{-3}S_{\mathrm{[CII]}}\Delta V\nu_{\mathrm{rest}}(1+z)^{-1}D_{\mathrm{L}}^2$, where $\nu_{\rm rest}$ is the [\ion{C}{2}] rest frequency (1900.5369 GHz) and $D_L$ is the luminosity distance in units of Mpc. We find $L_{\mathrm{[CII]}} = (1.46 \pm 0.22) \times 10^9\ L_{\mathrm{\odot}}$. 

We measured the [\ion{C}{2}] $158\ \micron$ line spectrum with a $1''.5$ diameter circular beam after subtracting the continuum emission. The line profile of [\ion{C}{2}] is clearly different from what was reported in \citet{2021ApJ...908..235I} (see Figure \ref{fig:spectra}a, \ref{fig:spectra}b), which is, as described in \S~\ref{sec:observation}, primarily due to the improved continuum subtraction in our analysis.

The resultant [\ion{C}{2}] spectrum is well represented by a single Gaussian function (Figure \ref{fig:spectra}b), and the fitting results are shown in Table 1.
The integrated value of this single Gaussian fit ($1.27 \pm 0.07\ \mathrm{mJy\ km\ s}^{-1}$) is consistent with the value measured in moment-0 map (Figure \ref{fig:two}b) within the error margin.
We derived the [\ion{C}{2}] redshift from this fit as $z_{\mathrm{[CII]}}=6.7224\pm 0.0003$ (Table \ref{tab:deluxesplit}). 
The \ion{Mg}{2}-based redshift is blueshifted 
relative to this \citep[$z_{\mathrm{MgII}}=6.699_{-0.001}^{+0.007}$][]{2019ApJ...880...77O}, indicating the presence of a nuclear ionized outflow of $-1040_{-50}^{+310}\ \mathrm{km\ s}^{-1}$ relative to the systemic velocity. \citet{2020ApJ...905...51S} investigated a sample of 38 luminous quasars and reported that the peak of the MgII-[CII] velocity shift distribution is located in the -560 to -340 km/s bin. Our target in this study is positioned at the high-velocity end of this distribution. This suggests the presence of a nuclear outflow in addition to the BAL. It is necessary to investigate whether the population of high-z red quasars occupies the high end of this distribution, and whether this outflow will impact the evolution of this galaxy in the future.

Starting with the discovery of the [CII] outflow in the high-z quasar J1148$+$5251 \citep[e.g.,][]{2012MNRAS.425L..66M, 2015A&A...574A..14C}, numerous [\ion{C}{2}] observations of high-z quasars have been conducted. However, outflows are rare, and the presence or absence of such outflows remains under debate, as observational and statistical analyses continue to yield differing conclusions \citep[e.g.,][]{2019A&A...630A..59B,2020ApJ...904..131N,2021ApJ...914...36I,2022ApJ...927..152M}.
As one of our prime motivations to study this red quasar is to explore signatures of AGN feedback on the galaxy-scale ISM and star formation, we attempted to identify broad wings in this [\ion{C}{2}] spectrum that might be due to fast AGN-driven outflows. 
To this end, we fit a double Gaussian function to the observed spectrum (Figure \ref{fig:spectra}c), but there is no improvement in $\chi^2 / \mathrm{d.o.f}$ (Table \ref{tab:deluxesplit}). 
Hence, there is no statistical evidence for the second component, and we cannot claim the presence of fast [\ion{C}{2}] outflows or the companion referred to by \citet{2021ApJ...908..235I}. 
This fact aligns with the results of the non-detection of a broad component in the deep stacking analysis of [\ion{C}{2}] emission in $27$ quasars at $z\gtrsim6$ \citep{2020ApJ...904..131N}. 

Following the standard prescription of [\ion{C}{2}] outflow studies \citep[e.g.,][]{2021ApJ...914...36I}, we here attempt to place rough upper limits on the properties of the neutral outflow, if one exists, by assuming that the outflow velocity is $v_{\mathrm{out}}\sim 500\ \mathrm{km\ s}^{-1}$. 
The spatial extent (radius) of the possible outflow is approximated as the half size of the 2D elliptical Gaussian fitted to the moment-0 map (see Table \ref{tab:deluxesplit}). Note that we do not consider any uncertainty of these assumptions for simplicity: our purpose here is to assess an order of magnitude impact of the possible AGN-driven outflow. 

We then created a moment-0 map by integrating over the range from $1400~\mathrm{km\ s}^{-1}$ to $2400~\mathrm{km\ s}^{-1}$ relative to the line center, and measured the rms in the areas free of the line emission. 
The resultant rms value is $0.029\ \mathrm{Jy\ km\ s}^{-1}$. We consider three times this rms value as the peak flux density of the potential outflow component, which corresponds to a $3\sigma$ upper limit of the [\ion{C}{2}] outflow luminosity of $<1.0\times10^{8}\ L_{\odot}$. From this, we derive the outflow mass ($M_{\mathrm{out}}$) and neutral outflow rate ($\dot{M}_{\mathrm{out}}$) as $M_{\mathrm{out}}<9.6\times10^7\ M_{\odot}$ and $\dot{M}_{\mathrm{out}}<33\ M_{\odot}\ \mathrm{yr}^{-1}$, respectively. This outflow rate reflects only the neutral atomic component, but \citet{2019MNRAS.483.4586F} argue that the molecular component also contributes a similar fraction by comparing the neutral mass outflow rate derived from the [\ion{C}{2}] line and \ion{Na}{1} D absorption line with the molecular mass outflow rate obtained from the CO transition line and OH absorption line in local ULIRGs.
Nevertheless, even if the total outflow rate ($\dot{M}_{\mathrm{out}}^{\mathrm{tot}}$) is twice the derived outflow rate above ($\dot{M}_{\mathrm{out}}^{\mathrm{tot}} = 66\ M_{\odot}~\mathrm{yr}^{-1}$), the resultant kinetic power of this outflow ($\dot{E}_{\mathrm{out}}^{\mathrm{tot}}=5.2\times10^{42}\ \mathrm{erg\ s}^{-1}$) corresponds only to 0.01\% of the bolometric luminosity of this quasar. 
This value is $\sim 500$ times smaller than what is expected in the blast energy-conserving feedback model \citep[e.g.,][]{2005Natur.433..604D,2012ApJ...745L..34Z}. 
Hence the AGN feedback of this quasar traced by the [\ion{C}{2}] line, even if it exists, will only have a limited impact on the host galaxy in terms of energy-conserving feedback.

\subsection{Dynamical Mass} \label{subsec;dyn}

\begin{figure}
    \centering
    \plotone{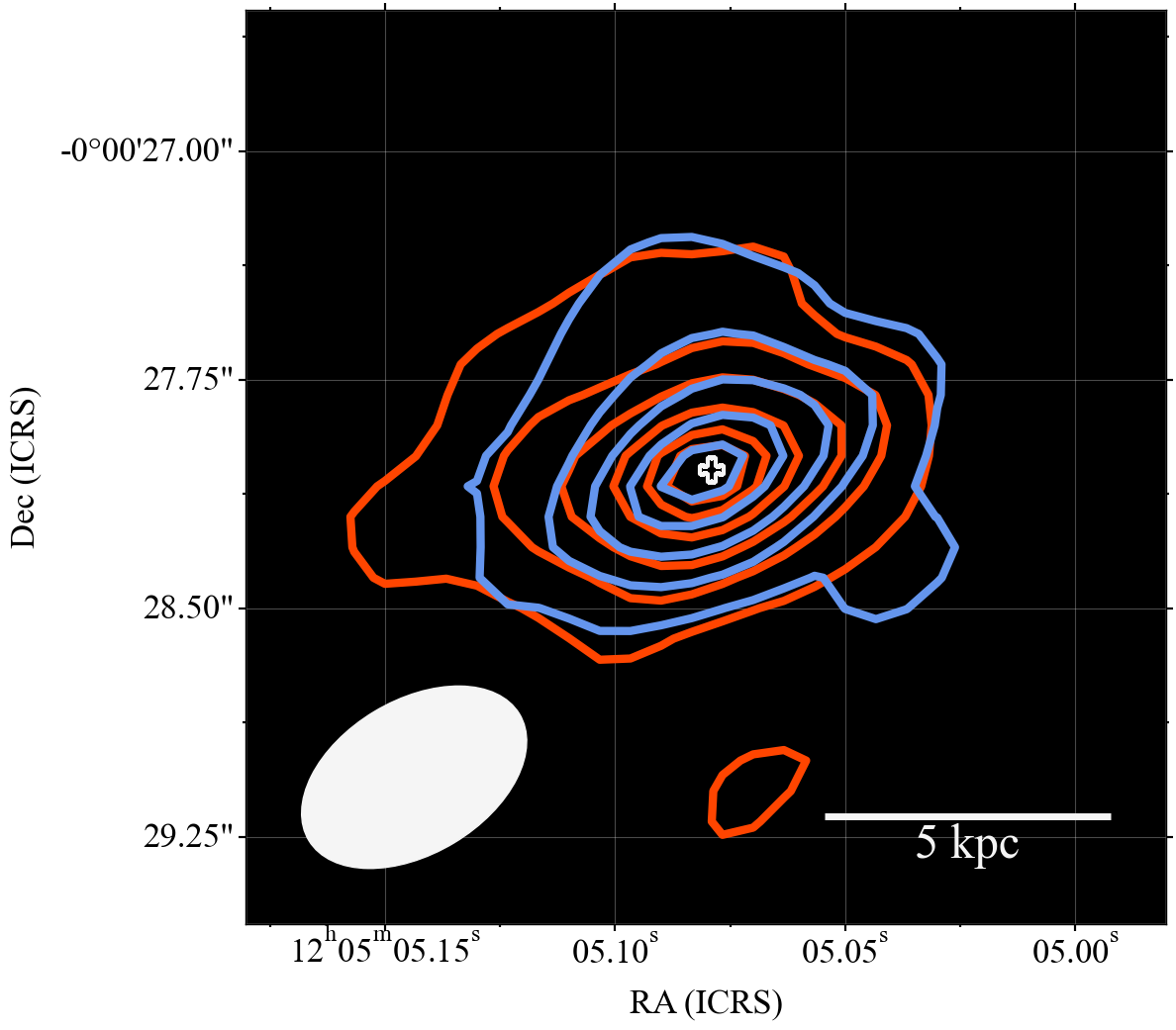}
    \caption{The velocity-integrated moment-0 map of the blue-shifted component ($-750\ \mathrm{km\ s}^{-1}$ to $0\ \mathrm{km\ s}^{-1}$) is shown as blue contours in this figure, while the red-shifted component ($0\ \mathrm{km\ s}^{-1}$ to $750\ \mathrm{km\ s}^{-1}$) is shown as red contours. The blue contours are plotted at levels of $3\sigma$, $6\sigma$, $9\sigma$, $12\sigma$, $14\sigma$ ($1\sigma = 0.036\ \mathrm{Jy\ beam}^{-1}\ \mathrm{km\ s}^{-1}$). For the red contours, the levels are $3\sigma$, $6\sigma$, $9\sigma$, $12\sigma$, $14\sigma$, $15\sigma$ ($1\sigma = 0.035\ \mathrm{Jy\ beam}^{-1}\ \mathrm{km\ s}^{-1}$). There is no evidence for an offset between the two, as would be expected if this were rotating. The black plus sign represents the center of the rest-frame UV light distribution observed by HSC \citep{Matsuoka2016}.}
    \label{fig:bluered}
\end{figure}

The [\ion{C}{2}] image and kinematics of this host galaxy are consistent with it being an elliptical galaxy supported by velocity dispersion. Based on this, we introduce the virial mass, given by $M_{\mathrm{dyn}}=3R\sigma^2/2G$, where $R$ is the galaxy radius, $\sigma$ is the velocity dispersion ($\sim \mathrm{FWHM_{[CII]}}/2.35$), and $G$ is the gravitational constant. Assuming $R=a_{\mathrm{maj}}/2$ using the beam-deconvolved major axis of the [\ion{C}{2}] emitting region estimated from a 2D Gaussian fit \citep[e.g.,][]{2018ApJ...854...97D, 2024arXiv240812299B}, we estimate $M_{\mathrm{dyn}}=(2.81\pm0.83)\times10^{10}~M_{\odot}$. 

To search for rotation, we created moment-0 maps for the blue-shifted and red-shifted components, corresponding to velocity ranges of $-750\ \mathrm{km\ s}^{-1}$ to $0\ \mathrm{km\ s}^{-1}$ and $0\ \mathrm{km\ s}^{-1}$ to $750\ \mathrm{km\ s}^{-1}$, respectively (Figure \ref{fig:bluered}). These maps reveal no spatial offset between the two components, indicating the absence of significant rotational motion on the spacial scales we can resolve. This finding further supports a dispersion-dominated nature.

We explored modeling the [\ion{C}{2}] emission as coming from a rotating disk. However, the current data have less than two beam elements across the [\ion{C}{2}] radius, insufficient for robust modeling of rotation using methods such as ${}^\mathrm{{3D}}\mathrm{BAROLO}$ \citep{2015MNRAS.451.3021D}. So we have tried to a less sophisticated approach here. The dynamical mass under the assumption that the [\ion{C}{2}] emission originates from a thin rotating disk is given by $M_{\mathrm{dyn}}/M_{\odot}=1.16\times 10^5\ (v_{\mathrm{circ}}/\mathrm{km\ s}^{-1})^2 (D/\mathrm{kpc})$ \citep[e.g.,][]{Wang2013,2015ApJ...807..180W,2018PASJ...70...36I}. The circular velocity, $v_{\mathrm{circ}}$, is approximated as $v_{\mathrm{circ}}=0.75\times \mathrm{FWHM_{\mathrm{[CII]}}}/\sin i$, where $i$ is the disk’s inclination angle. We determine $i$ as $i=\cos^{-1}(a_{\mathrm{min}}/a_{\mathrm{maj}})$, with $a_{\mathrm{min}}$ and $a_{\mathrm{maj}}$ representing the beam-deconvolved minor and major axes of the [\ion{C}{2}] emitting region, respectively. Here we assumed $D = 1.5 \times a_{\rm maj}$ to account for spatially extended fainter emission \citep[e.g.,][]{2010ApJ...714..699W,2019PASJ...71..111I}. 
We estimate $M_{\mathrm{dyn}}=1.3\times 10^{11}\ M_{\odot}$, which is a factor $\sim4$ higher than the $M_{\mathrm{dyn}}$ derived from the assumption of dispersion-dominated system.
Assuming that this $M_{\rm dyn}$ is dominated by stars \citep[as is usually adopted in $z > 6$ quasar studies, e.g.,][]{2015ApJ...807..180W,2016ApJ...816...37V,2019PASJ...71..111I}, we find that the host galaxy of J1205$-$0000 resides at the massive end of the $M_\star$ distribution for $z \sim 6-7$ galaxies \citep{2015A&A...575A..96G}, 
suggesting that this red quasar is one of the most evolved systems known at $z > 6$. Observations of higher spatial resolution are necessary to more precisely constrain the values \citep{2019ApJ...874L..30V,2019ApJ...880....2W}. 

In this paper, we advance the discussion by considering both possibilities. Specifically, we account for the possible existence of the rotational motion under the limit of spatial resolution, and both derived dynamical masses are comparable to the dynamical \citep[e.g.,][]{2021ApJ...911..141N} and stellar \citep[e.g.,][]{2023Natur.621...51D,2024ApJ...966..176Y,2024arXiv240907113O} masses of the luminous quasars.

\subsection{Star-Forming Activity} \label{subsec; SFR}
\begin{figure}[t]
\plotone{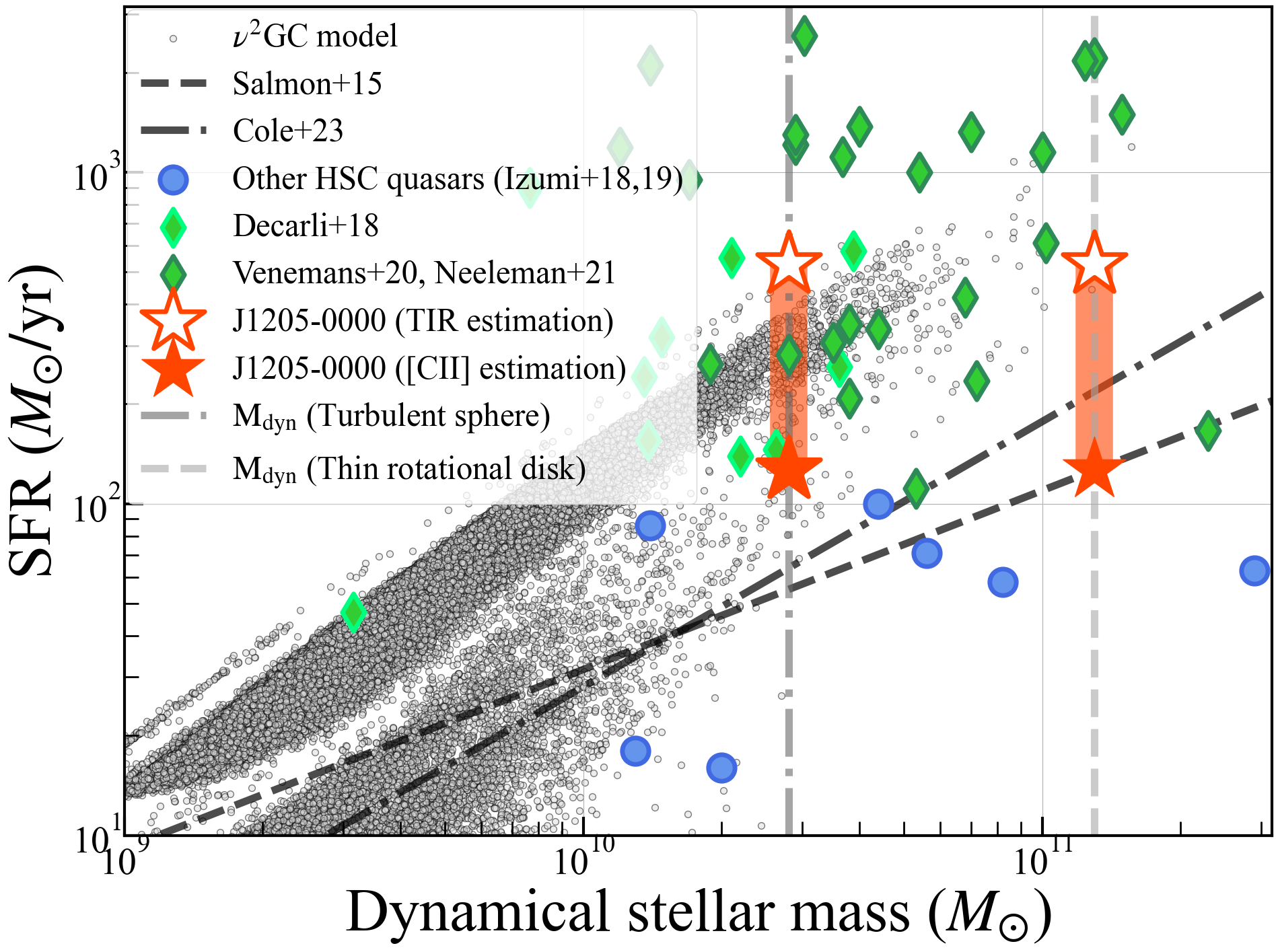}
\caption{Distribution of $M_{\rm dyn}$ (surrogate for $M_\star$) and TIR-based SFR of $z > 6$ quasar-host galaxies. The HSC quasars \citep[blue circles;][]{2018PASJ...70...36I,2019PASJ...71..111I}, optically luminous quasars \citep[green diamonds;][]{2018ApJ...854...97D, 2020ApJ...904..130V, 2021ApJ...911..141N}, and the results from the $\nu^2$GC semi-analytic model (gray dots; both the MS and the starburst sequences are evident), are shown. For reference, we also plot the observationally constrained MS of $z \sim 5-6$ galaxies \citep{Salmon2015}, and $6 < z < 7$ galaxies whose SFR is derived by averaging SFH over $10\ \mathrm{Myr}$ \citep{2023arXiv231210152C}. 
The open red star indicates J1205$-$0000 using the $L_{\rm TIR}$-based SFR, while the filled red star indicates the case with the $L_{\rm [CII]}$-based SFR. Note that the semi-analytic model and both MS lines use the stellar mass.}
\label{fig:SFR}
\end{figure}

We convert from $L_{\rm TIR}$ to SFR, using $\mathrm{SFR_{TIR}}=1.49\times 10^{-10}\ L_{\mathrm{TIR}}/L_{\odot}$ calibrated in the nearby galaxy NGC 6946 \citep{Murphy2011}, finding $\mathrm{SFR_{TIR}}=528\pm 8~M_{\odot}\ \mathrm{yr}^{-1}$ for J1205$-$0000. 
In order to discuss the level of star-forming activity of J1205$-$0000 relative to galaxies at comparable redshifts, 
we place J1205$-$0000 on the $M_\star$ vs SFR plane (Figure \ref{fig:SFR}). 
In this study, we will take $M_\star$ to be $M_{\mathrm{dyn}}$, corrected for the estimate of the gas mass. For the latter, we use the inferred dust mass, and assume a standard gas-to-dust mass ratio of 100 \citep[e.g.,][]{2007ApJ...663..866D}. We should account for the contribution of gas mass ($M_{\mathrm{gas}} = (1.78 \pm 0.03) \times 10^{10}\ M_{\odot}$) to the derived dynamical mass in \S~\ref{subsec;dyn}. Accordingly, the inferred stellar mass used ranges from $M_{\mathrm{dyn}} - M_{\mathrm{gas}}$, which is $(1.02 \pm 0.80) \times 10^{10}\ M_{\odot}$ for the dispersion-dominated case and approximately $11 \times 10^{10}\ M_{\odot}$ for the rotating disk assumption, up to the full value of $M_{\mathrm{dyn}}$ in both cases. However, due to uncertainties in the gas-to-dust mass ratio, dust temperature, and the precise structure of the system, the data shown as diamonds, circles, and stars in Figure \ref{fig:SFR} only display the dynamical mass.

The majority of normal star-forming galaxies form a sequence (main sequence = MS) on the $M_\star$--SFR plane, as seen from both at low redshift \citep[e.g.,][]{2007ApJ...670..156D} and at $z \sim 5-6$ \citep[e.g.,][]{2014ApJS..214...15S,Salmon2015} and $6<z<7$ \citep{2023arXiv231210152C}. 
Relative to this MS, we can discuss whether the host galaxy of J1205$-$0000 is starburst, normal, or quiescent. 

Figure \ref{fig:SFR} compares J1205$-$0000 (open star) with low-luminosity HSC quasars \citep{2018PASJ...70...36I,2019PASJ...71..111I}, optically luminous quasars \citep{2018ApJ...854...97D}, and predictions from the semi-analytic galaxy evolution model \citep[$\nu^2$GC model,][]{2016PASJ...68...25M,Shirakata2019}. 
The $\nu^2$GC model shows both the MS and a starburst sequence: the gap between these is artificial due to limited mass and time resolution. 
The MS constrained by observations at $z \sim 6$ \citep{Salmon2015} and at $6 < z < 7$ \citep{2023arXiv231210152C} are also shown as a reference. 

It is evident from Figure \ref{fig:SFR} that J1205$-$0000 is undergoing starburst activity comparable to the luminous quasars, while the low-luminosity quasars tend to lie on or even below the MS. This starburst nature is similar to what has been observed in optically luminous quasars \citep[e.g.,][]{2018ApJ...854...97D, 2020ApJ...904..130V, 2021ApJ...911..141N}.
Our result therefore suggests that both the starburst and the luminous quasar activity are coeval in both red and unobscured blue quasars, and AGN feedback in any form has not played a significant role in suppressing the star formation in them.

If we assume that the [\ion{C}{2}] emission primarily originates from star-forming regions, we can estimate the SFR of this quasar host galaxy by using the [\ion{C}{2}] line luminosity ($\mathrm{SFR_{\mathrm{[CII]}}}$) as well. 
With the relation calibrated in local \ion{H}{2}/starburst galaxies \citep[][]{derooze14}, namely $\log({\mathrm{SFR_{\mathrm{[CII]}}}}/M_{\odot}\ \mathrm{yr}^{-1})=-7.06+1.0\times \log(L_{\mathrm{[CII]}}/L_{\odot})$, we obtain $\mathrm{SFR_{\mathrm{[CII]}}} = 127 \pm 19\ M_{\odot}\ \mathrm{yr}^{-1}$, placing this object close to the MS (Figure \ref{fig:SFR}).
It has been confirmed by \citet{2020A&A...643A...3S} that the relation derived from local \ion{H}{2}/starburst galaxies is consistent with the SFR of galaxies at $z\sim4-8$.
This value is 4 times smaller than the $\mathrm{SFR_{TIR}}$. This discrepancy may indicate that the dust is partially heated by the AGN, boosting its strength \citep{2023MNRAS.523.4654T,2023MNRAS.518.3667D}. 

\section{Discussion} \label{sec:dis}
Our results show no evidence for strong [\ion{C}{2}] outflows in this red quasar (\S~\ref{subsec; line}). 
In addition, we found that the host galaxy is bright at FIR wavelengths, i.e., showing starburst-class (IR-based) or at least MS-class ([\ion{C}{2}]-based) star formation (\S~\ref{subsec; SFR}). 
These results suggest that putative negative AGN-feedback, at least at the levels expected in the energy-conserving feedback model \citep[e.g.,][]{2005Natur.433..604D,2012ApJ...745L..34Z}, has not played a significant role in quenching the star-forming activity of this quasar host galaxy.
This appears to contradict the galaxy evolution model that includes AGN feedback as a key process to regulate star formation (\S~\ref{sec:intro}). 
However, there is a room to circumvent this contradiction by considering (i) that J1205$-$0000 is in the very early stage of AGN feedback such that the feedback has not yet reached the scale of the host galaxy, or (ii) that there is significant AGN-driven outflow but it is not captured appropriately by [\ion{C}{2}] line emission.  These possibilities are detailed in the following. 

\subsection{Is the Outflow Too Young to Propagate to Host Galaxy Scales?}
A nuclear fast outflow is observed in the core region of J1205$-$0000, , giving rise to a BAL in \ion{N}{5} and \ion{C}{4}. Based on this, one simple explanation for the absence of the galaxy-scale outflow is that the nuclear fast outflow has not yet propagated to the scale of the host galaxy.

The timescale for outflow propagation to host galaxy scales can be estimated as $t=R_{\mathrm{host}}/v_{\mathrm{out}}$, where $R_{\mathrm{host}}$ is the radius of the host galaxy and $v_{\mathrm{out}}$ is the outflow velocity. We adopt $R_{\mathrm{host}}=1.5~\mathrm{kpc}$, which is approximately the radius of the FIR continuum (dust) emitting region of J1205$-$0000.
We derive upper and lower limits of the timescale by adopting the two scenarios for the outflow velocity; one where the nuclear fast outflow (with $v_{\mathrm{out}}=4000~\mathrm{km/s}$ observed by \citet{2019ApJ...880...77O}) propagates directly to the scale of the host galaxy, and the other where we use the velocity of the [\ion{C}{2}] outflow at $z=7.07$ (with $v_{\mathrm{out}}=500~\mathrm{km/s}$ observed by \citet{2021ApJ...914...36I}). The range of resultant timescales is $t\sim(0.4-3)~\mathrm{Myr}$. Given that the typical duration of the quasar phase at $6\lesssim z\lesssim 7$ is $\sim10~\mathrm{Myr}$ \citep[e.g.,][]{2017ApJ...840...24E,2023MNRAS.522.4918S,2024ApJ...969..162D}, it is possible that J1205$-$0000 represents a short-term transitional event, accounting for only $\sim(4-30)~\%$ of its quasar lifetime. This result is consistent with the estimated blowout timescale of $10^5~\mathrm{yr}$ for outflows in red quasars at $z<3$ \citep[e.g.,][]{2020MNRAS.495.2652L, 2021ApJ...906...21J, 2022MNRAS.517.3377S}, and is also roughly consistent with the existence of red quasars, which account for $\sim 10~\%$ of the type 1 SDSS quasars at $z<2.2$ \citep[e.g.,][]{2023ApJ...954..156K, 2024ApJ...972..171K}. In summary, in this interpretation, we are witnessing the onset of AGN activity ($< 4-30~\%$ of the total quasar lifetime) for the case of J1205$-$0000, and significant AGN-driven feedback onto the host galaxy would happen at a later evolutionary stage.


\subsection{Is the Outflow Driven by Radiation pressure?}


We consider models of radiation pressure-driven outflows \citep[e.g.,][]{2005ApJ...618..569M, 2015MNRAS.451...93I, 2016MNRAS.463.1291I, 2018MNRAS.479.2079C, 2018MNRAS.473.4197C}, which do not include the propagation of BALs. 
As red quasars are expected in a transition phase from a dusty starburst, this radiation pressure-driven outflow would be reasonable for them. 
For example, \citet{2018MNRAS.479.2079C} predicts that radiation pressure-driven outflows of $\dot{E}_{\rm out}/L_{\rm Bol} \lesssim 0.1\%$, consistent with the upper limit value of $0.01\%$ in J1205$-$0000 (\S~\ref{subsec; line}), can still have a significant impact on terminating host galaxy-scale star formation \footnote{Note that \citet{2018MNRAS.479.2079C} modeled a much more luminous quasar than J1205$-$0000, hence it is unclear whether their results are applicable for J1205$-$0000}. However, since the derived $\dot{M}_{\mathrm{out}}^{\mathrm{tot}}$ is $\sim9$ times smaller than $\mathrm{SFR_{TIR}}$, it is not dominant as a feedback process.



\subsection{Dense outflow}
Another simple interpretation for the absence of massive [\ion{C}{2}] outflows in J1205$-$0000 is that [\ion{C}{2}] is not an appropriate tracer of outflows. 
A remarkable example is the luminous quasar J2054$-$0005 at $z = 6.04$, which does not show any clear evidence of [\ion{C}{2}] outflows \citep{Wang2013}. 
However, recent OH $119\ \micron$ absorption line observations toward this quasar clearly show deep blueshifted components, indicative of the presence of massive ($\sim 1500~M_\odot$ yr$^{-1}$) and fast ($\sim 670$ and $\sim 1500$ km s$^{-1}$) molecular outflows, to which [\ion{C}{2}] observations are insensitive \citep{Salak2024}. Similarly, \citet{2023ApJ...944..134B} reported similar findings in a different quasar. 
In addition, \cite{Spilker2020} reported clear detection of host galaxy-scale molecular outflows using the OH $119\ \micron$ absorption line in $z > 4$ starburst galaxies, which again lack fast [\ion{C}{2}] components. 
These results strongly suggest that the [\ion{C}{2}] line may not be a reliable tracer of neutral outflows in extreme objects such as starburst galaxies and quasars at $z>4$. 
Although the reason for the absence of [\ion{C}{2}] outflows but for the presence of molecular outflows remains unclear, one possible scenario is that the bulk outflow component of these systems is denser than the critical density of [\ion{C}{2}] ($n_{\mathrm{crit}}=2.8\times10^3\ \mathrm{cm}^{-3}$), leading to a suppression of [\ion{C}{2}] line emission by collisional de-excitation.
The majority of the mass at the centers of local galaxies is in the form of molecular gas \citep[e.g.,][]{1995A&A...304....1H}, and multi-line and multi-transition analyses of circumnuclear disks (on scales of $\sim100~\mathrm{pc}$) in nearby AGNs often report high gas densities of $\sim 10^{4-5}~\mathrm{cm}^{-3}$ \citep[e.g.,][]{2014A&A...570A..28V, 2020ApJ...898...75I}. Therefore, it is highly likely that the central region around a quasar is generally at high density. Additionally, the outflow propagating through the host galaxy ISM may compress the surrounding gas, potentially leading to a selectively high-density environment.
Observations of dense gas tracers such as [\ion{O}{1}] $63\ \micron$ ($n_{\mathrm{crit}}=4.7\times10^5\ \mathrm{cm}^{-3}$), [\ion{O}{1}] $145\ \micron$ ($n_{\mathrm{crit}}=9.4\times10^4\ \mathrm{cm}^{-3}$) and OH $119\ \micron$ ($n_{\mathrm{crit}} \gtrsim 10^9\ \mathrm{cm}^{-3}$) in J1205$-$0000 will provide crucial data to test this scenario.

\subsection{Too slow outflow}
\citet{2024MNRAS.533.1733W} suggest that, even if there is a fast ($5000-10000~\mathrm{km s}^{-1}$) nuclear outflow, cold outflows (e.g., traced by [\ion{C}{2}] emission) at the host galaxy scale could have low velocities ($\lesssim 100~\mathrm{km s}^{-1}$), hence the outflow properties might be significantly underestimated.
For example, for the case of a quasar (AGN luminosity of $10^{45}~\mathrm{erg~s}^{-1}$) simulated in \citet{2024MNRAS.533.1733W}, the mass outflow rate estimated from the high-velocity component of cold outflows ($>100~\mathrm{km/s}$) is a factor of $\sim8$ smaller than the outflow rate including slow ($\sim10-100~\mathrm{km~s}^{-1}$) velocity components. Distinguishing such a slow outflow observationally would be difficult because, for our usual estimation of outflow properties, we naively adopt values estimated from high-velocity (i.e., broad wing) components, beyond the $\mathrm{FWHM_{[CII]}}$ that would represent the rotation of the host galaxy. Quasars at $z > 5.7$ generally have $\mathrm{FWHM_{[CII]}}/2 \gtrsim 100~\mathrm{km/s}$ \citep[e.g.,][]{Wang2013, 2018ApJ...854...97D, 2019PASJ...71..111I, 2020ApJ...904..130V}, making it challenging to separate outflow from non-outflow gas components according to this model prediction.

\begin{figure}[t]
\plotone{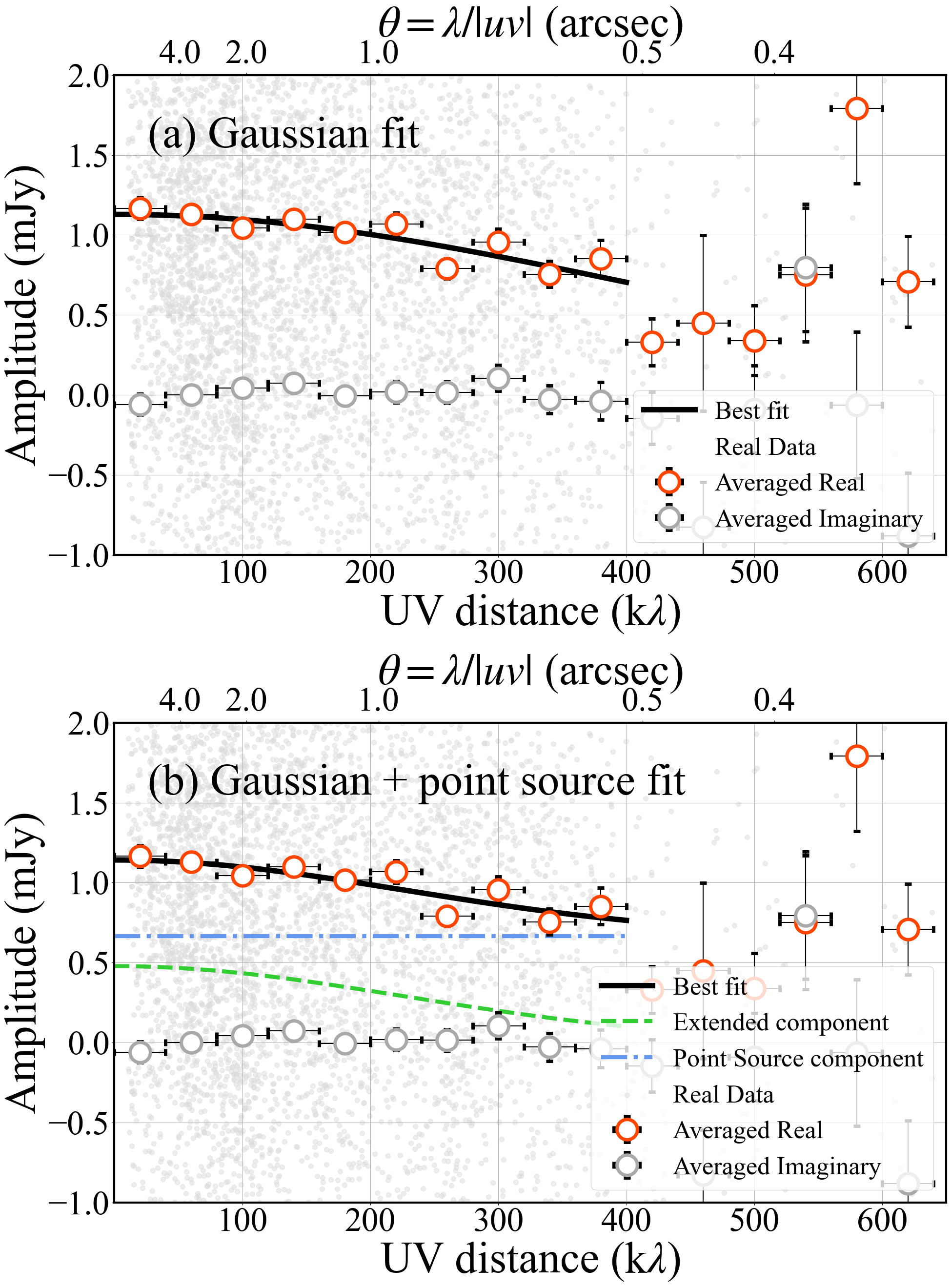}
\caption{The real and the imaginary components of visibility data of the continuum emission as a function of $uv$ distance, binned over $40\ \mathrm{k\lambda}$. (a) A circular Gaussian model fit to the binned real part of visibility, resulting in the goodness of fit of $\chi^2/\mathrm{d.o.f}=1.37$. (b) Point source fitting in addition to a circular Gaussian to separate a point-source component and extended component. This plot is slightly worse ($\chi^2/\mathrm{d.o.f}=1.50$) than the fit of (a). These fits are to the average real amplitudes from $0\ \mathrm{k\lambda}$ to $400\ \mathrm{k\lambda}$. The imaginary amplitudes are distributed around zero.
\label{fig:uvamp}}
\end{figure}

\subsection{Caveat: actual star-forming activity}
If the above scenarios in which massive outflows in J1205$-$0000 are actually present are correct, we have to explain the currently observed behavior that many of the (luminous) quasar host-galaxies are starburst systems, i.e., there is no signature of quenching of star formation, in most quasars (but see \citet{2024arXiv240907113O}), including J1205$-$0000 in Figure \ref{fig:SFR}.

It is noteworthy in this context that recent ALMA observations toward $z > 4$ quasars have begun to identify point source components, in addition to extended components (i.e., host galaxies) \citep[e.g.,][]{2021ApJ...914...36I,2023MNRAS.523.4654T}. 
These works have attributed the point sources to unresolved contamination from the central AGN/quasar \citep{Salak2024}. 
That is, unresolved observations of rest-frame IR continuum emission in (intrinsically) luminous quasars might overestimate the IR luminosity driven by star formation, and thus also overestimate the SFR. 

We thus attempt to assess the FIR flux contamination from the AGN in J1205$-$0000. We will work in the $uv$ plane. This analysis of the complex visibility data allows us to study structures that are unresolved in the image plane. An unresolved point source (a delta function in the image plane) appears as a constant (flat) component in the $uv$-distance vs $uv$-amplitude plane and is regarded as AGN contamination here. We note that due to the angular resolution of the data, this component may also include nuclear starburst activity \citep[e.g.,][]{2021MNRAS.506.3946D}. Figures \ref{fig:uvamp}a and \ref{fig:uvamp}b show the circular Gaussian fitting (single component fit) and the circular Gaussian plus constant function fitting (double component fit) to the $uv$-distance vs $uv$-amplitude data of J1205$-$0000 (bin-width; $40\ \mathrm{k\lambda}$, fitting range; $0\ \mathrm{k\lambda}\sim400\ \mathrm{k\lambda}$), created using the {\tt\string CASA} task {\tt\string plotms}, after shifting the phase center to the FIR continuum peak coordinate. Our single component fit finds a FWHM of $0''.18 \pm 0''.02$, whereas the double component fit given FWHM of $0''.34\pm 0''.19$. Both of the FWHMs are roughly comparable to the beam-deconvolved size of FIR continuum emission measured on the image plane (Table \ref{tab:deluxesplit}). 

With the double component fit, we measure the total flux density of the extended Gaussian component as $0.47\pm0.31\ \mathrm{mJy}$ and that of the point source component as $0.64\pm0.32\ \mathrm{mJy}$. 
If this is the case, more than half of the continuum emission originates from the compact component that is likely heated by the central (intrinsically) luminous quasar, and the SFR taking only this extended component into account becomes $212 \pm 138~M_{\odot}~\mathrm{yr}^{-1}$. 
However, largely because of the fact that our data is noisy at $\gtrsim 400\ \mathrm{k}\lambda$ (the scales that are most sensitive to the point source component), we did not see any significant improvement of the double component fit ($\chi^2/\mathrm{d.o.f}=1.50$) over the single component fit ($\chi^2/\mathrm{d.o.f}=1.34$). 
Hence we conclude that the addition of the point source is not statistically supported at this moment.  
Higher resolution observations that are more sensitive to the emission of $\gtrsim 400\ \mathrm{k}\lambda$ component are necessary to further test this scenario. 

In any of the three scenarios described above, however, the expected AGN feedback on star formation is weak since the star formation activity of J1205$-$0000 remains above the MS.
This holds even after removing potential contamination of the quasar nucleus to the IR continuum emission (Figure \ref{fig:uvamp}).
In summary, it is still too early to have a decisive conclusion on the genuine absence or presence of cold outflows solely from the current [\ion{C}{2}] line data, given the unconstrained reliability of the line as a tracer of cold outflows. 
Future high sensitivity and high resolution observations of the OH absorption line, [\ion{O}{1}] line, and associated continuum emission, will provide a more robust view on the feedback in this intriguing phase of SMBH evolution.

\section{Conclusions}\label{sec:conclusion}
We have re-analyzed the [\ion{C}{2}] 158\micron\ line and the underlying rest-frame FIR continuum emission data of the red quasar J1205$-$0000 at $z=6.72$ obtained by ALMA. This red quasar was discovered in the HSC-SSP optical survey \citep{Aihara2018} using the Subaru telescope. J1205$-$0000 appears as a low-luminosity quasar ($M_{\rm 1450} = -24.4\ \mathrm{mag}$), but once the dust extinction is corrected for, it is considerably more luminous, $M_{\rm 1450} = -26.1\ \mathrm{mag}$.

We improved the continuum subtraction in the ALMA spectroscopy from \citet{2021ApJ...908..235I}, significantly changing the results. They can be summarized as follows.


\begin{enumerate}
    \item We detected significant [\ion{C}{2}] line and FIR continuum emission. The [\ion{C}{2}] line emission is resolved ($\sim5\ \mathrm{kpc}$), and is found to be more extended than the FIR continuum emission, consistent with what \citet{2021ApJ...908..235I} found. The [\ion{C}{2}] line luminosity is $L_{\mathrm{[CII]}} = (1.46 \pm 0.22) \times 10^9\ L_{\mathrm{\odot}}$ and the total IR continuum luminosity is $L_{\mathrm{TIR}}=(3.54\pm 0.05)\times 10^{12} L_{\mathrm{\odot}}$ (assuming $T_{\mathrm{dust}}=47\ \mathrm{K},\ \beta=1.6$). 

    \item Unlike \citet{2021ApJ...908..235I}, the [\ion{C}{2}] line profile is well modelled by a single Gaussian with $\mathrm{FWHM_{\mathrm{[CII]}}}=404\pm 27~\mathrm{km\ s}^{-1}$. We found no statistical improvement in the fitting by adding a broad line component (i.e., double Gaussian fit). Hence we did not detect the presence of fast [\ion{C}{2}] outflows in our data. Our crudely estimated $3\sigma$ upper limit on the mass outflow rate, after accounting for the other phase gas flows, is $\lesssim 70~M_\odot$ yr$^{-1}$. 

    \item We assume a velocity dispersion-supported structure, resulting in $M_{\mathrm{dyn}} = (2.81 \pm 0.83) \times 10^{10}\ M_{\odot}$. Additionally, considering the possibility that J1205$-$0000 is a rotating system, we derive the dynamical mass based on the assumption of a rotating geometrically thin disk. Using the FWHM of the [\ion{C}{2}] line profile ($404 \pm 27\ \mathrm{km\ s}^{-1}$) and its spatial extent, we estimate the dynamical mass of J1205$-$0000 to be $M_{\mathrm{dyn}} \sim 1 \times 10^{11}\ M_{\odot}$. Similar to other optically luminous quasars at comparable redshifts, the host galaxy of J1205$-$0000 in this interpretation ranks among the most massive galaxies known at $z \sim 6-7$.
    
    \item The SFRs estimated from the total IR continuum luminosity, and from the [CII] line luminosity, are $\mathrm{SFR_{TIR}}=528\pm 8~M_{\odot}\ \mathrm{yr}^{-1}$ and $\mathrm{SFR_{\mathrm{[CII]}}} = 127 \pm 19\ M_{\odot}\ \mathrm{yr}^{-1}$, respectively. Assuming $M_{\mathrm{dyn}}=M_{\mathrm{*}}$ and using the TIR-based SFR, we find that J1205$-$0000 is a starburst galaxy, lying above the observed star-forming main-sequence at that redshift. However, our results depend on the method to determine $M_{\mathrm{dyn}}$; different assumptions put the galaxy on the star-forming MS. These SFRs may be overestimated due to potential contamination from the AGN in the rest-frame FIR emission.

    \item Even though J1205$-$0000 shows BALs in its rest-frame UV spectrum, our $\dot{E}_{\rm out}/L_{\rm Bol}~(< 0.01\%)$ is considerably smaller than the energy-conserving feedback model that assumes propagation of nuclear BAL winds to the scale of the host galaxy. Although further detailed assessment is required, one can circumvent this inconsistency if the AGN feedback of J1205$-$0000 is still in a very early stage, i.e., the nuclear outflow has not yet reached the host galaxy. Alternatively, feedback models which invoke radiation pressure-driven outflows only require $\dot{E}_{\rm out}/L_{\rm Bol} \lesssim 0.1\%$, which is consistent with our upper limit.
    
    \item Considering the recent detections of dense molecular outflows probed by OH 119 \micron\ absorption in quasars at $z > 6$, including the case of a luminous quasar without significant [\ion{C}{2}] outflows, we speculate that the average gas density of the outflows in J1205$-$0000 is greater than the critical density of [\ion{C}{2}] ($n_{\mathrm{crit}}=2.8\times10^3\ \mathrm{cm}^{-3}$), causing the [\ion{C}{2}] line emission to be very weak due to collisional de-excitation. There is also a possibility that the outflow properties are underestimated because the velocity of cool outflows, as traced by [\ion{C}{2}], is quite low.
\end{enumerate}

Based on our results, we cannot decisively conclude whether an AGN-driven massive cool outflow is present or not in this red quasar (and in other high-z quasars as well) without studying gas flows in multiple phases. 
Future multiphase gas observations of this and similar quasars, particularly dense gas tracers such as OH 119 \micron\ and [\ion{O}{1}] 63, 145 \micron\ lines ($n_{\mathrm{crit}}=4.7\times10^5, 9.4\times10^4\ \mathrm{cm}^{-3}$, respectively), will shed light on the presence or prevalence of massive AGN-driven outflows at $z > 6$. High resolution continuum emission observations will resolve the structure either in the image plane or in the $uv$ plane (Figure \ref{fig:uvamp}), which will allow a more accurate assessment of the AGN contamination to the currently observed SFR.

\begin{acknowledgments}
This paper makes use of the following ALMA data: ADS/JAO.ALMA\#2019.1.00074.S. 
ALMA is a partnership of ESO (representing its member states), NSF (USA) and
NINS (Japan), together with NRC (Canada), NSC and ASIAA
(Taiwan), and KASI (Republic of Korea), in cooperation with
the Republic of Chile. The Joint ALMA Observatory is
operated by ESO, AUI/NRAO and NAOJ. 
Data analysis was carried out on the Multi-wavelength Data Analysis System operated by the Astronomy Data Center (ADC), National Astronomical Observatory of Japan (NAOJ). 

T.H., S.B., Y.M., T.K., K.K., M.O. and T.I. were supported by the Japan Society for the Promotion of Science (JSPS) KAKENHI grant No. [22H01258, 23K22529], 23H05441, 21H04494, JP23K13153, [22H04939, 23K20035, 24H00004], 24K22894 and [21H04496, 23K03462] respectively.
T.K. is grateful for support from the RIKEN Special Postdoctoral Researcher Program.
K.I. acknowledges support under the grant PID2022-136827NB-C44 provided by MCIN/AEI/10.13039/501100011033 / FEDER, UE. M.S. was supported by the ALMA Japan Research Grant of NAOJ ALMA Project, NAOJ-ALMA-339.
\end{acknowledgments}

%

\vspace{5mm}
\facilities{ALMA}


\software{astropy \citep{2013A&A...558A..33A,2018AJ....156..123A}, {\tt\string CASA} \citep{McMullin2007,2022PASP..134k4501C}}



\bibliography{sample631}{}
\bibliographystyle{aasjournal}



\end{document}